\documentclass[12pt]{scrartcl}
\usepackage{graphicx} 
\usepackage{amsmath} 
\usepackage[a4paper, margin=3.0cm]{geometry}
\usepackage{authblk}
\usepackage{float}
\usepackage{hyperref}

\title{Quantification of Electron Energy-Loss Spectra}

\author[1,2]{Pavel Potapov}
\author[3]{Giulio Guzzinati}

\affil[1]{IFW-Dresden, Dresden, Saxony, Germany}
\affil[2]{TEMDM, Dresden, Saxony, Germany }
\affil[3]{CEOS GmbH, Heidelberg, Baden-Württemberg, Germany}

\begin{document}

\maketitle

\noindent
\textit{This manuscript summarizes the recent developments in EELS  quantification flow as will be implemented in the CEOS Panta Rhei and TEMDM software. This should serve as a technical reference for the algorithms used in the software. }

\section{Motivation}

The methods for quantifying Electron Energy-Loss Spectra (EELS) have been intensively developed over the past half-century, as summarized most comprehensively in Egerton’s monograph \cite{Egerton}. In particular, the accuracy of quantification improved significantly through the model-based fitting of spectral profiles \cite{Verbeeck2004}. However, recent advances in instrumentation introduce new challenges. First, the large size of modern EELS spectrum-images necessitates automated quantification procedures with little or no user interaction. Second, EELS data are now available over wide energy ranges — up to several thousands eV — where the common assumptions of standard quantification approaches may no longer hold.

In particular:
\begin{itemize}

	\item Within a sufficiently narrow energy region, the EELS background can be reasonably modeled by a power law: $A E^{-r}$ where $E$ in energy-loss and $A$, $r$ are constants. However, this simple power-law form becomes inaccurate over larger ranges extending to several thousand eV.

	\item Solid-state effects can significantly modify the shapes of observed EELS edges, causing them to deviate from theoretically calculated profiles. Such effects are still not comprehensively accounted for in existing quantification methods.

	\item Plural scattering can dramatically distort the shapes of EELS edges compared with those calculated under the assumption of single-scattering events.

\end{itemize}

These issues, along with several practical aspects of EELS quantification, are addressed in the present manuscript.

\section{Fitting Ranges}

The choice of fitting ranges for the quantification of experimental EELS spectra is not straightforward. One common approach is to use the entire recorded energy-loss range \cite{Verbeeck2004}. However, in many cases, large portions of the spectrum do not contain features relevant for quantification. Including such regions in the fit may only increase the likelihood of artifacts due to limitations of the fitting model.

Another strategy is to define fitting ranges for all known ionization edges. This would require compiling and maintaining an extensive reference list, which we consider overly complex.

The problem becomes even more nuanced when the fitting procedure requires the definition of \textit{pre-edge} regions — energy intervals preceding the edge onset and used to extrapolate the background. In our view, additional transitional zones between the edges and pre-edge regions are also necessary. We refer to these zones as the \textit{margin} regions. Their role is to provide a buffer between drastically different ranges.

One of the first comprehensive theoretical calculations \cite{Leapman1980} demonstrated that the strength of EELS cross-sections decreases sharply with increasing the ionization threshold (onset) while their shape becomes more prolonged towards higher energies (Fig. 4, 7, 8, 12 in \cite{Leapman1980}). On the other hand, the exponential decay of EELS intensity with increasing energy-loss amplifies noise and necessitates longer fitting intervals at higher energies to ensure reliable fit.

Based on these considerations, we define all fitting ranges fixed  in logarithmic energy scale. This approach implies that the absolute width of the fitting range increases with the energy onset of the edge. In our current implementation, we fix the edge fitting width at \textbf{0.6} log(eV), pre-edge width at \textbf{0.1} log(eV)  and margin width at \textbf{0.015} log(eV). The resulted ranges for  selected edges are listed in Table 1.


\begin{table}[ht!]
	\centering
	\caption{Fitting Energy Ranges (in eV) for Selected Ionization Edges}
	\begin{tabular}{lcccc}

		Edge         & Pre-Edge Start & Margin Start & Edge Start & Edge End \\
		\hline
		Si $L_{2,3}$ & 89             & 97           & 99         & 180      \\
		C $K$        & 256            & 279          & 284        & 517      \\
		N $K$        & 362            & 395          & 401        & 730      \\
		Ti $L_{3}$   & 412            & 449          & 456        & 830      \\
		Ti $L_{2}$   & 418            & 455          & 462        & 841      \\
		O $K$        & 481            & 524          & 532        & 969      \\
		Cu $L_{3}$   & 842            & 917          & 931        & 1696     \\
		Cu $L_{2}$   & 860            & 936          & 951        & 1732     \\
		Ge $L_{3}$   & 1101           & 1198         & 1217       & 2217     \\
		Ge $L_{2}$   & 1129           & 1229         & 1248       & 2274     \\
		Si $K$       & 1664           & 1812         & 1839       & 3351     \\
		Sr $L_{3}$   & 1755           & 1911         & 1940       & 3534     \\
		Sr $L_{2}$   & 1816           & 1977         & 2007       & 3656     \\
	\end{tabular}
	\label{tab:range_fitting}
\end{table}

As shown in the table, the fitting ranges are generally broad and often overlap when multiple ionization edges are present in a spectrum. In such cases, fitting ranges  are merged into a single  patch, and the corresponding edges are fitted simultaneously. However, the pre-edge regions are not merged. Instead, the lowest pre-edge region is retained. The reason behind this will be explained in the following section.

\begin{figure}[ht!]
	\caption{ Illustration of fitting ranges in logarithmic scale. Edges with higher onset energies correspond to wider fitting ranges. Overlapping edges are merged into a single fitting patch, with only the lowest pre-edge region retained for background modeling.}
	\includegraphics[width=1.0\textwidth]{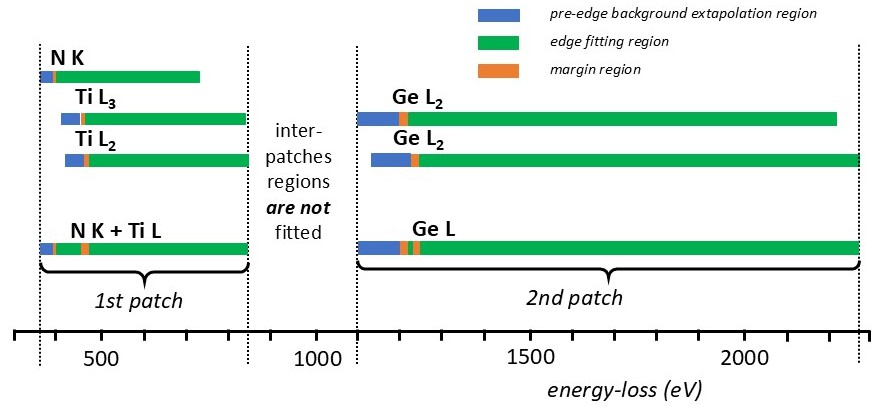}
	\label{fig1}
\end{figure}

Fig.~\ref{fig1} illustrates this fitting strategy using the example of the N $K$, Ti $L$, and Ge $L$ edges. In this case, the algorithm constructs two separate fitting patches, each containing its own pre-edge region. No fitting is performed at energies outside these two patches.

\section{Background Subtraction}

EELS ionization edges are typically superimposed on a significant background, which often greatly exceeds the intensity of the edges themselves. This is in contrast to Energy-Dispersive X-ray Spectroscopy (EDX), where ionization lines  usually significantly exceed the background level. In EELS, background evaluation may represent the largest source of error in spectral analysis. Therefore, accurate background modeling and its separation from the ionization edges is, in our view, a central challenge in EELS quantification.

The background in EELS can arise from multiple sources, including single-electron excitations, tails of low-energy collective excitations, multiple scattering \textit{etc}. In most cases, it is not feasible to identify the background sources and model them accordingly, therefore, empirical modeling is  employed. It has been shown that, within energy ranges of a few hundred eV, the background  can be reasonably approximated by a Power Law of the form $B(E) = a E^{-r}$ \cite{Egerton}.
The parameters of Power Law can be easily deduced from the energy region preceding ionization edges by solving

\begin{equation}
	log(B(E)) \approx log(a) - r \cdot log(E)
	\tag{1}
\end{equation}

The Power Law can be then extrapolated to the entire required energy range. However, the extrapolation to the energy ranges approaching or exceeding 1000 eV typically fails, suggesting that  a simple Power Law is not able to approximate background in such wide ranges, at least with a fixed power law exponent $r$.

\begin{figure}[h!]
	\caption{(a) Power law exponent \textit{r} for Cu samples of varying thicknesses, calculated over 100 eV-wide energy intervals not consisting of any characteristic edges. Sample thickness ranges from 0.1 to 1.6 in terms of mean free path. (b) Power Law exponent for different materials measured in wider energy range. The drop in the 500–600 eV region of Cu samples is due to eventual residual O-K edge contributions. All measurements were performed with a  125 mrad collection semi-angle }
	\includegraphics[width=\textwidth]{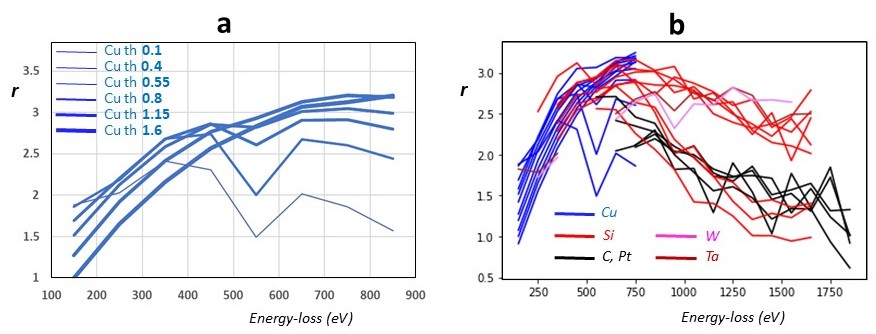}
	\label{fig2}
\end{figure}

The behavior of the exponent \textit{r} as a function of energy loss is not well understood and may depend on a number of factors. Fig.~\ref{fig2}a shows the variation of \textit{r} for copper samples of different thicknesses, while Figure \ref{fig2}b compares measurements across different materials using the same instrument and collection semi-angle. Notably, \textit{r} may increase or decrease with energy loss. These trends might arise from intrinsic material properties or instrumental effects, such as varying the effective collection semi-angle with energy loss. Still, the evolution of \textit{r} appears to follow a smooth trend, often resembling a quadratic or cubic polynomial.

This uncertainty has motivated several authors to go beyond the simple Power Law model. Cueva et al. \cite{Cueva} proposed a linear combination of two Power Law terms, while Van den Broeck et al. \cite{Wauter2025} utilized up to five terms, including those with negative magnitudes. The latter  required quadratic programming to avoid the non-monotonicity and convexity issues.

\subsection{Smooth Background}

In this paper, we propose an alternative method: explicit modeling a smooth change in the Power Law exponent with energy-loss. We refer to the method as the “smooth background” model. This approach introduces non-linearity into the fitting procedure, which, however, can be efficiently managed as we show below. Importantly, the gain in accuracy for background modeling may justify the increased complexity of the model.

\begin{figure}[h!]
	\caption{Power law in standard (a) and logarithmic (b) coordinates. In log space, the exponent \textit{r} corresponds to the slope of a straight line. Deviations from this straight line can be modeled with a cubic polynomial. Parameters $y_0, r_0$ and $y_{fin}, r_{fin}$ define the  boundary conditions similar to that for a cubic spline.}
	\centering
	\includegraphics[width=1.0\textwidth]{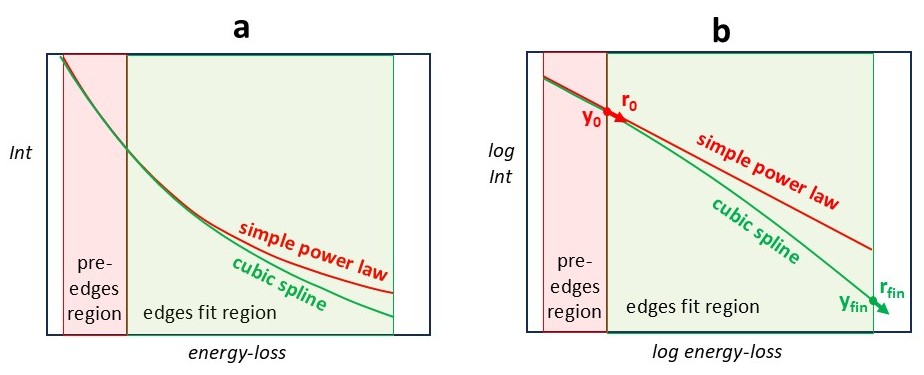}
	\label{fig3}
\end{figure}

To explain our method, recall that  a Power Law (Fig.~\ref{fig3}a) appears as a straight line with a slope equal to the exponent \textit{r}  in logarithmic coordinates (Fig.~\ref{fig3}b). If this slope changes smoothly, we arrive at a scenario where the exponent varies continuously with energy-loss as shown by a green curve in the figure. The simplest way to achieve this is by replacing the straight line with a cubic polynomial.

We define the polynomial parameters using two boundaries: initial and final points - each with their values and derivatives - similar to that for a cubic spline. The initial point lies in the pre-edge region, where power law parameters are reliably calculated; the final point is found by optimal fitting. This introduces only two additional fitting parameters — $y_{fin}$ and $r_{fin}$ — similar to the approach of Cueva et al. \cite{Cueva}. However, the task is non-linear as discussed below.

One key advantage of our method is that the modeled background, by definition, exactly matches the measured background in the pre-edge region — something not guaranteed by the models of Cueva and Van den Broeck. Another benefit is the method’s ability to closely reproduce the natural, smooth evolution of the exponent \textit{r} within its quadratic variations. However, we acknowledge that particularly irregular \textit{r} variations may not be captured by our smooth background approach.

Although our method is empirical, it must satisfy certain physical constraints, such as a monotonic decrease  with increasing energy loss. While no fundamental law prohibits non-monotonic EELS background, we are not aware of any experimental observations reporting it. Another common expectation is background smoothness — \textit{i.e.}, the absence of pronounced inflection points or shoulders. Some authors interpret this as the convexity requirement (positive second derivative everywhere \cite{Wauter2025} ). In our view, this condition is overly restrictive. Strict convexity cannot be maintained in regions where \textit{r}  increases with energy loss. This is evident in the green curve in Fig.~\ref{fig3}b, where it is always possible to find two points with the curve lying above the chord connecting them, which means breakdown of convexity. Simple analysis confirms that this loss of convexity persists when transforming back to standard coordinates. Yet, an increase in \textit{r}  with energy-loss is frequently observed, as demostrated in Fig.~\ref{fig2}.

Appendix A derives conditions to ensure the monotonicity of the modeled background and provides constraints for the cubic spline to ensure this. These constraints also promote background smoothness, even if they do not always ensure strict convexity. In addition, we impose a physically motivated constraint: the modeled background must not exceed the experimental spectrum at any energy loss.

\begin{figure}[h!]
	\caption{Comparison of smooth background and simple Power Law methods. The two approaches give same results at the start of the fitting range but may diverge significantly toward the end. (b) shows a clear advantage of the smooth background model, where the simple Power Law fails by crossing the experimental spectrum.}
	\includegraphics[width=\textwidth]{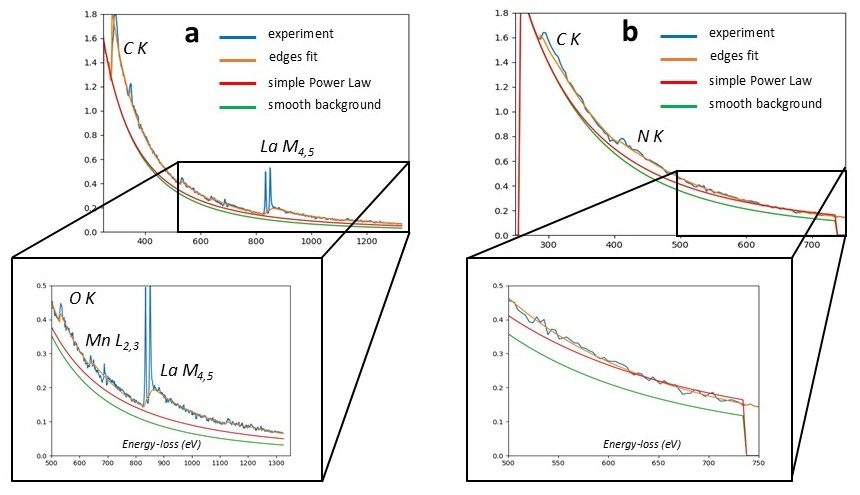}
	\label{fig4}
\end{figure}

In many cases, the smooth background closely follows the extrapolation of a simple Power Law. However, discrepancies between the two methods become pronounced as the fitting range extends (Fig.~\ref{fig4}) and/or when the signal-to-background ratio is small. The advantage of our method is especially clear when the simple Power Law eventually crosses the experimental spectrum, whereas the smooth background remains consistent, as demonstrated in Fig.~\ref{fig4}b.

\subsection{Pre-edges region}

As stated in the previous subsection, the smooth background algorithm inherits the initial power-law parameters $r_0$ and $y_0$ from the pre-edge region. These values are not included in the fitting flow and therefore must be determined accurately from Eq.~(1). This typically does not pose a problem, except in the following cases:
\begin{itemize}
	\item The pre-edge region contains a noticeable excursion, such as afterglow traces of the zero-loss peak or defective pixels.
	\item The spectrum is so noisy that the Power Law parameters cannot be reliably extracted.
\end{itemize}

When the pre-edge region is manually controlled, the user can easily detect these issues and adjust the region by shifting or extending it. However, since we target fully automatic processing, the following algorithm is applied.

First, the Power Law evaluation is performed within the predefined pre-edge region. The region is then shifted toward lower energies by half of its width, and the evaluation is repeated. If the value of $r_0$ is reproduced with sufficient precision (currently, we use the criterion $r = r_0 \pm 0.2$), the parameters are accepted. Otherwise, the shift is repeated until reproducibility is achieved. The lower bound for such ``crab-like'' steps is set either to a fixed energy loss of 50~eV (commonly recognized as the upper energy limit for plasmon excitations) or to the upper boundary of the preceding quantification patch window, if it exists.

If this lower bound is reached while reproducibility is still not achieved, this typically indicates that the Power Law parameters fluctuate strongly due to excessive noise. In this case, the pre-edge region is extended to include all crab-like steps of shifting ranges. The maximum number of steps is currently limited to 6. This implies that the pre-edges region may be extended by up to a factor of 4, statistically reducing the noise amplitude by  a factor of two.

\section{Theoretical Cross-Sections}

Within the first Born non-relativistic approximation, the theoretical EELS double differential cross-section as a function of scattering vector  $q$ and energy loss $E$ is expressed as

\begin{equation}
	\frac{d^2\sigma(E,q)}{dEdq} = \frac{e^4}{4\pi\epsilon_0^2m_ev^2Eq} \frac{df(E,q)}{dE}
	\tag{1}
\end{equation}

where $e$, $m_e$ and $v$ are charge, mass and velocity of the incident electron, $\epsilon_0$ is a vacuum permittivity and $f(q,E)$ is the generalized oscillation strength (GOS) for the corresponding EELS ionization edge.

In our quantification procedure,  we used the GOS values calculated by Segger et al. \cite{Segger} for  perturbation of a single atom. Within this approach, the atomic wavefunctions were computed self-consistently with a modified version  of the  program of Hamman \cite{Hamann} under the local density approximation using the exchange–correlation potential of Perdew and Zunger \cite{Perdew} (see \cite{Segger} for further references and an in depth explanation). The resulting GOS values were tabulated for the full set of EELS edges, and are freely available at \cite{zenodo}.

Alternatively, relativistic cross-sections can be employed \cite{Zhang2025}. To our knowledge, this makes the only slight difference for typical quantification cases.

To match observable spectra, the theoretical cross-sections must be integrated over the experimental angular range, i.e. within the  collection semi-angle $\beta$. Accordingly, the tabulated GOS values were first linearly interpolated over the experimental angular range and then integrated in two-dimensional rings using Simpson’s composite rule.

When the convergence semi-angle $\alpha$ is nonzero, $\beta$ must be replaced by the effective collection semi-angle calculated according to (4.72) in \cite{Egerton}. Alternatively, a more sophisticated approach presented in \cite{KOHL1985265} can be used, but is not yet included within our software.

\begin{figure}[ht!]
	\caption{Computational flow of the "smooth background" algorithm. The outer loop involves non-linear fitting of background parameters $y_{fin}$ and $r_{fin}$. The inner loop is a linear fit  (consequent or parallel) of reference cross-sections to the residual signal.}
	\includegraphics[width=0.8\textwidth]{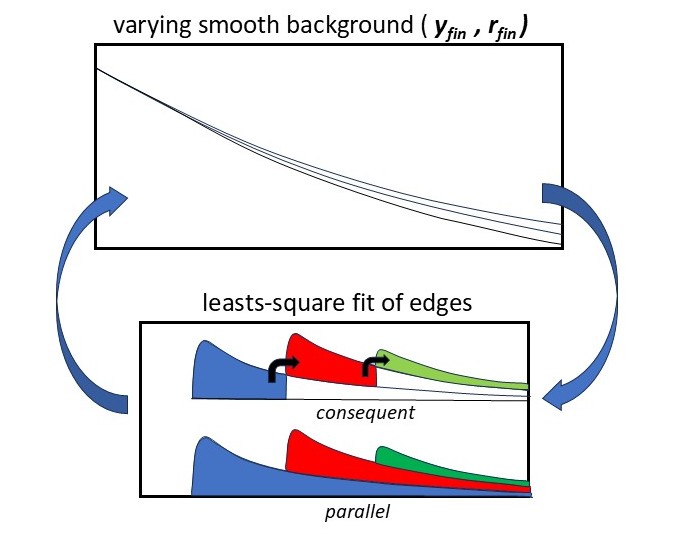}
	\label{fig11}
\end{figure}

\section{Fitting Procedure}

The fitting procedure consists of two nested loops — an outer loop and an inner loop — as shown in Fig.~\ref{fig11}.

The outer loop optimizes the background parameters for the entire fitting patch, which may contain several ionization edges. The fitting begins with a simple Power Law background ($r_{fin} = r_0$) estimated from the pre-edge region of the patch. The parameters $y_{fin}$ and $r_{fin}$ are then iteratively adjusted to minimize the final loss function.

In the inner loop, the background is subtracted, and the remaining signal is fitted using the profiles of all edges assumed to be present within the  patch. This fitting can be performed in two ways: \textit{consequent} or \textit{parallel}, as illustrated at the bottom of Fig.~\ref{fig11}.

The \textit{consequent} method fits the edges one after another, progressing from lower to higher onset energies. The lowest-energy edge is first fitted by least squares within an energy range where it does not overlap with any other edge. If the resulting edge weight is negative, it is set to zero to preserve the physical interpretation of the fit. The fitted edge profile is then extended over the entire fitting range and subtracted from the residual signal. The procedure is repeated for the next edge in energy until all edges have been considered.

The \textit{parallel} method fits all edges within the patch simultaneously using a non-negative least-squares (NNLS) algorithm. To enforce non-negative contributions from individual edges, NNLS employs an additional nested iterative procedure.

Both approaches have advantages and limitations. Consequent fitting is often more accurate for low-energy edges but may lead to error accumulation at higher energies. It is potentially faster because its computational complexity scales as $O(N^2)$, where $N$ is the number of fitting channels. Thus, dividing the full fitting range into smaller segments reduces the computational cost quadratically. In practice, however, the consequent approach involves substantial overhead operations and is often slower than the highly optimized matrix-inversion routines employed by the NNLS fitter. On the other hand, the NNLS approach is incompatible with the fine-structure weighting scheme discussed in Section 6.1.

After completion of the inner loop, the final loss function is evaluated as the sum of squared differences between the fitted model and the experimental data. The outer loop then updates $y_{fin}$ and $r_{fin}$, and the entire procedure is repeated until convergence.

\begin{figure}[ht!]
	\caption{Typical loss function landscape for the "smooth background" optimization problem. Both $\Delta$ and $k$ parameters are constrained within narrow bounds. As a rule, the loss function is smooth and convex.}
	\includegraphics[width=1.\textwidth]{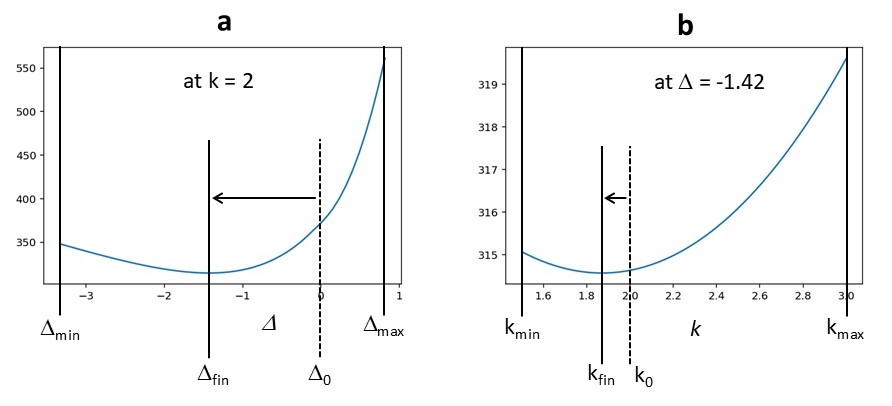}
	\label{fig12}
\end{figure}

The outer-loop optimization shown in Fig.~\ref{fig11} (adjustment of $y_{fin}$ and $r_{fin}$) involves exponential transformations and therefore renders the overall fitting procedure non-linear. Compared with linear methods, non-linear optimization presents two principal challenges: the possible existence of local minima and increased computational cost. The first issue is mitigated by imposing narrow bounds on the fitting parameters, as described in Appendix A. Fig.~\ref{fig12} shows a typical loss-function landscape plotted as a function of the two dimensionless fitting parameters $\Delta$ and $k$ (see Appendix A for their definitions). The loss surface is generally smooth and exhibits a single well-defined minimum.

The computational cost of the proposed method is indeed higher than that of purely linear fitting approaches. Nevertheless, because the loss function is typically convex (Fig.~\ref{fig12}), the optimization problem can be solved efficiently using modern non-linear optimization algorithms. By default, we employ the bounded limited-memory Broyden--Fletcher--Goldfarb--Shanno (L-BFGS-B) algorithm \cite{scipyBFGS}. This quasi-Newton method estimates gradients numerically and converges rapidly to the minimum of the convex loss function, typically requiring only a small number of iterations.

\section{Weighted Fitting}

After accurately subtracting the background, the remaining signal is assumed to arise from a linear combination of characteristic EELS ionization edges. The optimal combination is determined \textit{via} least-squares fitting, minimizing the squared difference between the model and the experimental spectrum over the \textit{entire} fitted energy range.

However, different regions of the spectrum might not contribute equally to the overall quality of the result. Some energy regions carry higher informational value and require stricter matching, while others may tolerate larger deviations. To address this, we introduce a weighting function $\omega(E)$, ranging from 0 to 1. This function assigns relative importance to different energy regions during the fitting process.

The least-squares fitting then minimizes the (squared) difference between the weighted model and the weighted experimental signal, according to:

\begin{equation}
	\min_{p_i} \int \left[ \omega(E) \cdot S(E) - \omega(E) \cdot \sum_i p_i \, \sigma_i(E) \right]^2 \, dE
	\tag{2}
\end{equation}

where $S(E)$ is the experimental spectrum after background subtraction, $\sigma_i(E)$ are the calculated edges cross-sections with their contributions $p_i$.

As shown later, introducing $\omega(E)$ allows solving various issues related to the optimal fitting.

\subsection{Account for Fine Structure}

Even with perfect background subtraction, accurate fitting between experimental signal and theory is not guaranteed. This is because the theoretical edge profiles are calculated for isolated atoms, while actual spectra are significantly influenced by solid state effects. Fortunately, these effects are largely localized near the edge onset and diminish with increasing energy-loss.

The simplest method to handle this issue is to exclude from the fit a fixed region immediately following the onset. This is implemented in \cite{GatanQuant}, where the user can interactively define the width of the excluded region to achieve the best visual agreement between the remaining spectrum and the atomic cross-section. However, this manual approach is incompatible with our goal of minimal user interaction.

An alternative strategy is to introduce additional fitting functions specifically tailored to model solid-state effects in the near-onset region \cite{Verbeeck2006}. This approach is effectively similar to the exclusion \cite{GatanQuant} unless the additional functions are constrained by some \emph{a priori} assumptions. Yet, reasonable criteria for such constraints are not clear, although recent studies \cite{Jannis2025} have made progress toward understanding this problem. Anyway, this approach significantly increases the number of fitting parameters.

\begin{figure}[h!]
	\caption{Experimental (a) almost noise-free and (b) noisy Ti $L_{2,3}$ edge after background subtraction compared to the atomic cross-section. Best match was achieved by manual interactive exclusion of the near-onset region. Residual deviations reveal solid state effects.}
	\includegraphics[width=1.0\textwidth]{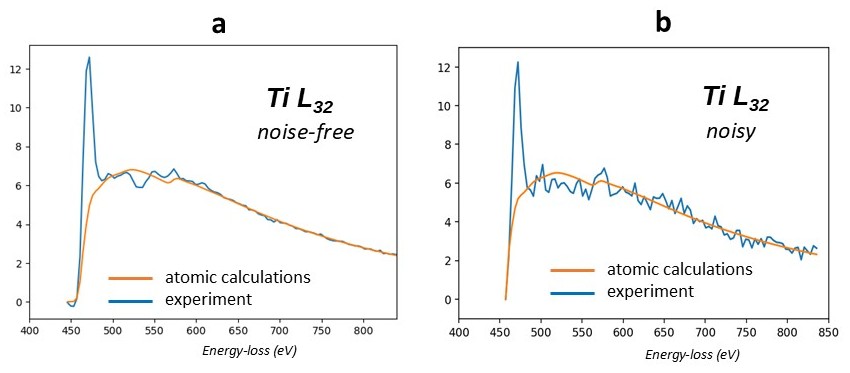}
	\label{fig5}
\end{figure}

We propose a different, purely empirical, solution, based on experimentally observed deviations from theoretical atomic cross-sections. For this, we selected several elements known to exhibit most pronounced fine structures. Other elements with more moderate fine structures are expected to fall naturally within the proposed framework.

Fig.~\ref{fig5}a shows an example for the Ti $L_{2,3}$ edge. After accurate background subtraction, we manually, iteratively excluded an initial energy-loss region to achieve the best fit between experiment and theory in the remaining spectral range. The deviations in the near-onset region were then extracted and, where applicable, deconvolved in order to separate $L_2$ and $L_3$ (or $M_4$ and $M_5$) contributions. The $L_3/L_2$ and $M_4/M_5$ ratios were determined using the maxima of the corresponding white lines. Finally, all measurements were normalized to the maximum of the theoretical cross-section.

\begin{figure}[p]
	\centering
	\caption{Normalized deviations from atomic cross-sections for (a) $K$, (b) $L$, and (c) $M$ edges in selected elements. For (b) and (c), deconvolution of $L_{2,3}$ and $M_{4,5}$ splitting was applied, though without sub-pixel precision, resulting in artifacts (arrows). Most deviations fall within the empirically derived Gaussian envelopes (shaded regions).}
	\includegraphics[width=0.6\textwidth]{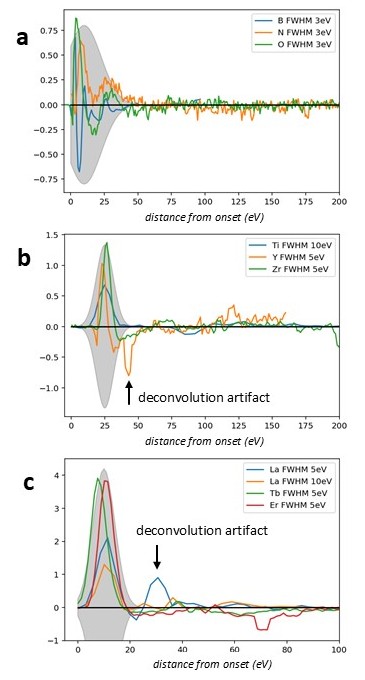}
	\label{fig6}
\end{figure}

Fig.~\ref{fig6} summarizes normalized deviations for $K$, $L$, and $M$ edges in selected elements. This visual analysis demonstrates that most deviations fall within well-defined limits, which can be approximated by an analytical function. In the preliminary release of this approach \cite{QuantMM2025}, we used an exponentially decaying function, but further testing and analysis on more data reveals that a Gaussian function is more appropriate for this purpose as illustrated in Fig.~\ref{fig6}.   It is important to note that Fig.~\ref{fig6} displays the \textit{maximum} expected deviations; actual deviations in other elements are likely to lie within these shaded areas.

\begin{table}[h!]
	\centering
	\caption{Parameters of the empirical limiting envelop for fine structure deviations}
	\begin{tabular}{lccc}

		       & Magnitude & Middle (eV) & $\sigma$ (eV) \\
		\hline
		K edge & 0.8       & 10          & 25            \\
		L edge & 1.5       & 25          & 20            \\
		M edge & 4.0       & 10          & 12            \\
	\end{tabular}
	\label{tab:elnes}
\end{table}

The experimental fine-structure deviations vary drastically between edge types. As summarized in Table 2, the amplitude of deviations for M edges can exceed those for K edges by a factor of  five. We have not yet analyzed N edges.

To account for these fine-structure deviations in the fitting, we construct a weighting function in the following way. Experimental spectra  inevitably differ from theory due to both solid-state effects and random noise as in Fig.~\ref{fig5}b. Assuming statistical independence of these effects, their variances add up. This leads to the following relationship:

\begin{equation} \label{eq3}
	\omega^2(\Delta E) \cdot \left( \Delta^2(\Delta E) + \sigma_N^2 \right) = \sigma_N^2
	\tag{3}
\end{equation}

where $\Delta E$ is the energy-loss above the edge onset, $\omega(\Delta E)$ is the weighting function, $\Delta(\Delta E)$ is the deviation due to solid state effects and $\sigma_N^2$ is the noise variance assumed to be constant within the considered energy range. This equation implies that in regions far above the onset, where the fine structure is relaxed, $(\Delta(\Delta E) \approx 0$, $\omega(\Delta E)=1$. Based on this, the weighting function is derived as:

\begin{equation} \label{eq4}
	\omega^2(\Delta E) = \frac{1}{1 + \frac{\Delta^2(\Delta E)}{\sigma_N^2}}
	\tag{4}
\end{equation}

Such weighting strategy is sensitive to the noise level in experimental spectra, which is a desirable feature. Noisy spectra anyway exhibit huge variance around the base-line even when the model agrees well with the underlying signal (Fig.~\ref{fig5}b).  Because least-squares fitting minimizes squared deviations, the account for fine structure is less important in this case.

\begin{figure}[H]
	\caption{(a) Normalized empirical deviations as in Fig.~\ref{fig6} and calculated weighting functions for fictive, illustrative (b) $K$, (c) $L$, and (d) $M$ edges. The weighting depends on the level of noise observed in experimental spectra although this dependence is weak.   A 25 eV splitting was assumed for both $L_{2,3}$ and $M_{4,5}$.}
	\includegraphics[width=\textwidth]{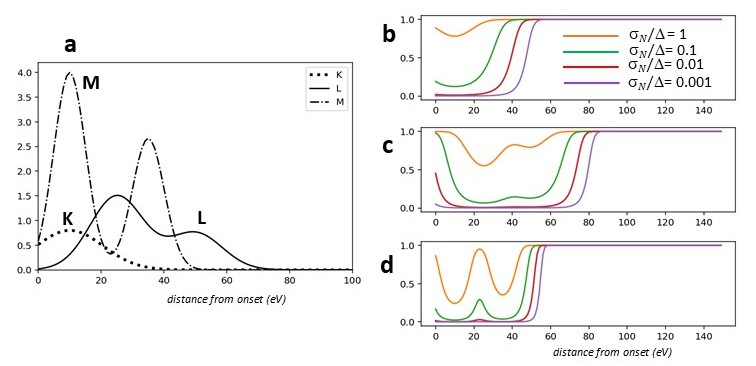}
	\label{fig7}
\end{figure}

Fig.~\ref{fig7} illustrates calculated weighting functions for $K$, $L$, and $M$ edges. As noted above, the weighting depends on the estimated noise level, which we currently normalize from the mean background-subtracted signal — a rough but sufficient approach. As shown in Fig.~\ref{fig7}, the weighting function responds rather to the order of magnitude, not to the exact value of the noise. To ensure numerical stability, we apply a lower cutoff: $\sigma_N / \Delta_{max} \geq 0.001$.

\begin{figure}[h!]
	\caption{Fit of La oxide spectrum before (a) and after (b) fine structure correction. While the O $K$ edge is only mildly affected, the La $M$ fit is improved substantially.}
	\includegraphics[width=1.0\textwidth]{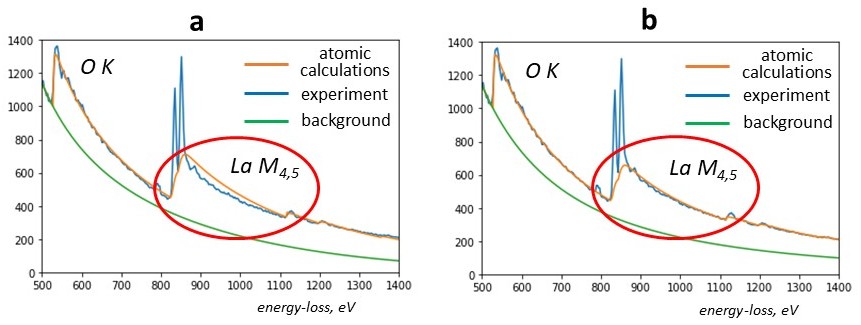}
	\label{fig8}
\end{figure}

Fig.~\ref{fig8} presents an example of fine structure correction. The correction has a minor impact on the O $K$ edge fit but significantly improves fitting of the La $M_{4,5}$ edge. Before correction, the experimental spectrum exceeds the theoretical prediction up to 300 eV above the onset, followed by a gradual underestimation likely to continue beyond the recorded range. After correction, the model and experiment agree over the entire observed energy window, apart from the near-edge white-line structure.

As evident from Fig.~\ref{fig7}, our method is effectively equivalent to excluding the near-onset region from the fit. However, unlike the hard-edge approach of \cite{GatanQuant}, our method introduces a kind of soft exclusion boundary, determined not arbitrarily but based on experimental observations. Another advantage is a built-in handling of edge splitting (e.g. $L_{2,3}$ and $M_{4,5}$ ), which is difficult to manage using fixed exclusion windows. Finally, our approach allows for relaxation of the exclusion requirements in high-noise situations, where information is insufficient for fitting.

One limitation of the proposed method compared to \cite{Verbeeck2006} is the requirement of consequent fitting when several edges overlap in the same energy region. Implementation of our approach in parallel fitting currently poses significant challenges.

\subsection{Poisson Weighting}

\begin{figure}[ht!]
	\caption{(a) Spectrum of a Si–O–C sample where large deviations around 300 eV dominate the fit and suppress more relevant differences at $\sim$900 eV, resulting in an inaccurate C/O ratio. (b) Application of Poisson weighting corrects the imbalance.}
	\includegraphics[width=1.0\textwidth]{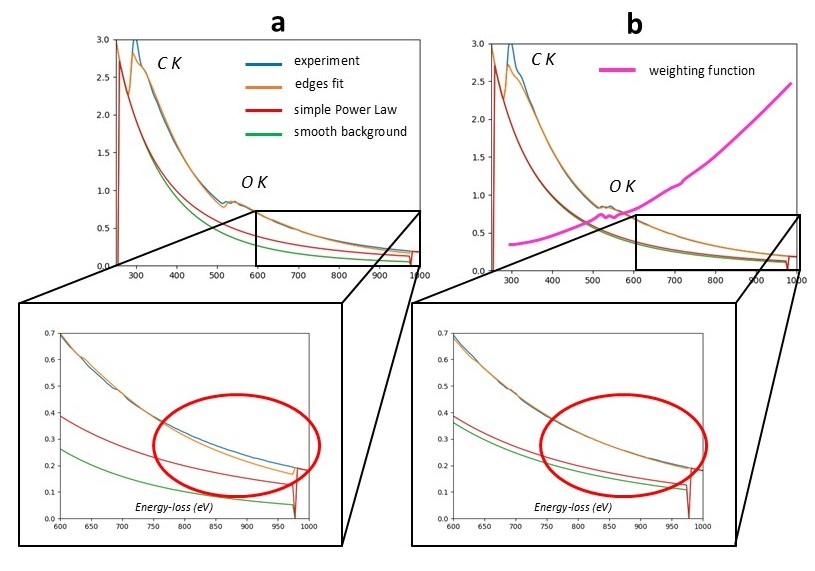}
	\label{fig9}
\end{figure}

When a fitting patch spans a wide energy range, the spectral intensity can vary dramatically across it, which may introduce inaccuracy in the fitting process. The dominant noise component in experimental EELS spectra is Poisson noise, whose variance scales with the mean intensity. However, standard least-squares fitting procedure minimizes the total squared deviation without accounting for the varying noise levels across the energy-loss axis.

Similarly, non-noise deviations from the calculated atomic cross-sections are not uniformly distributed across the spectrum. The absolute values of the atomic cross-sections decrease  with increasing energy-loss, meaning that their  deviations at low energy-losses might overshadow smaller,  but more critical, deviations at higher energy-losses. This can degrade the accuracy of the fit and distort elemental quantification.

Fig.~\ref{fig9}a illustrates this issue in a spectrum from a Si–O–C sample. The deviations between  experiment and theory at $\sim$300 eV are large in the absolute value. This dominates the overall fit and masks more relevant deviations occurring at $\sim$900 eV, thereby distorting the C/O ratio. 

The weighting function that solves the problem is the reciprocal of the spectrum \( S(E) \) before background subtraction:

\begin{equation} \label{eq5}
	\omega(E) =
	\begin{cases}
		\frac{1}{S(E)}, & S(E) > 0 \\
		0,              & S(E) = 0
	\end{cases}
	\tag{5}
\end{equation}

This weighting function is applied to the background subtracted spectrum $S(E)$, effectively resolving the fitting imbalance between low-loss and high-loss regions (Fig.~\ref{fig9}b).

\begin{figure}[ht!]
	\caption{(a) Poisson weighting function  exhibits dips at the places of white line in La-Mn oxide spectrum that (b) automatically accounts for the fine-structure deviations from the atomic cross-section.}
	\includegraphics[width=0.6\textwidth]{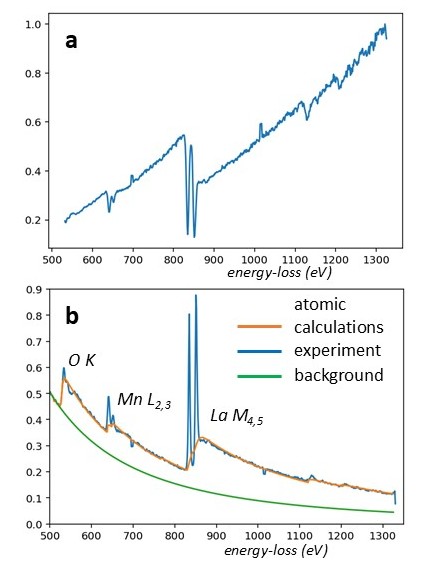}
	\label{fig_12}
\end{figure}

Interestingly, Poisson weighting also helps to mitigate the impact of fine structure variations. Fig.~\ref{fig_12} demonstrates that the La $M_{4,5}$ and Mn $L_{2,3}$ edges are fitted quite reasonably despite not using the fine structure correction described in the previous section.  This occurs because the most prominent fine structure features in $L$ and $M$ edges are white lines, i.e. huge intensity peaks. These peaks result in local dips in the weighting function, effectively down-weighting the regions most affected by fine structure. Although this is an incidental effect, that works only for \textit{positive} deviations, nevertheless it solves the issue of white lines with nearly same efficiency as the more complex treatment of Section 6.1.

\subsection{Other Weighting}

This weighting strategy can be extended to other situations where specific energy regions should contribute less to the fit. These include: (i) transitional zones between edge onsets and background/preceding edges (margin regions), (ii) pixels affected by afterglow effects in CCD detectors, and (iii) virtual pixels located in module gaps of hybrid  cameras.

\newpage
\section{Account for Plural Scattering}

Theoretical scattering cross-sections for specific elemental edges are typically calculated under the assumption of a single inelastic scattering event. This assumption holds reasonably well only for very thin TEM samples. In more realistic scenarios, an incident electron may undergo multiple successive inelastic interactions — a phenomenon known as \textit{plural scattering}. While the probability of double core-loss scattering is negligible, the likelihood of an electron experiencing a core-loss event followed by one or more low-energy-loss events is high in EELS core-loss spectroscopy.

The probability of plural scattering increases with increasing the sample thickness. To describe this process, it is convenient to introduce a dimensionless (relative) thickness $th = t/\lambda$, where $t$ is an actual sample thickness and $\lambda$ is a mean free path of an electron in a given material measured in the same units as $t$. Although $\lambda$ accounts for all inelastic events, it is effectively dominated by low-loss processes (e.g., plasmons), which occur much more frequently than high-loss events.

As a result, the spectrum of the core-loss signal is redistributed with adding the plasmon-like tails at higher energies.
This was not a major concern in the early days of EELS quantification, when the total integrated core-loss intensity was used to estimate elemental concentrations. Because the plural scattering enhancement was roughly similar for different edges, the resulting elemental ratios remained reasonably accurate.

However, this situation changed significantly with the advent of edge fitting techniques, where the experimental core-loss edge is compared to a reference profile. Since plural scattering distorts the spectral shape, it must now be properly accounted for in quantitative analysis.

One common approach is to deconvolve the experimental core-loss spectrum using the low-loss one, followed by comparison with theoretical single-scattering cross-sections. This procedure, often implemented \textit{via} the Fourier-ratio method, however tends to amplify high-frequency noise. As an alternative,  theoretical cross-sections are convolved with experimental low-loss spectra and then compared to the experimental core-loss spectra \cite{Verbeeck2009}. We adopt this convolution-based method in our quantification procedure.

\begin{figure}[ht!]
	\caption{Experimental (a) low-loss and (b)  background subtracted Cu $L_{2,3}$ core-loss spectra of samples with varying relative  thickness $th$.  (c) shows corresponding theoretical cross-section convolved with the experimental low-loss spectra. The energy resolution (FWHM) is 10 eV. Low-loss spectra are normalized to the height of a zero-loss peak while core-loss spectra are normalized to the total area. For easier comparison with the other edges, the energy-loss in (b, c) is counted from the Cu $L_{23}$ onset (931eV). }
	\includegraphics[width=\textwidth]{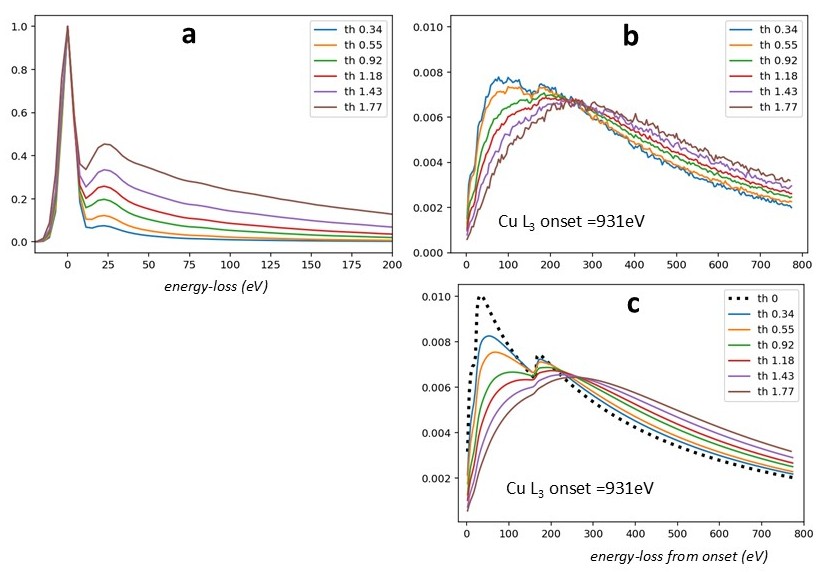}
	\label{fig_13}
\end{figure}

\subsection{Convolution with Available Low-Loss Spectra}

Fig.~\ref{fig_13}
presents the  treatment of a pure Cu sample with relative thicknesses $th$ ranging from 0.3 to 1.8. The thickness was determined from the low-loss spectra (Fig.~\ref{fig13}a)
by taking a logarithm of the ratio of the total intensity to the zero-loss one (equation (3.95) in \cite{Egerton}). The Cu $L_{2,3}$ profiles  change quite significantly with increasing the thickness (Fig.~\ref{fig_13}b),
i.e. at $th$ $>$ 1.5, their shape has actually little common with the case of $th \approx0.3$. However, these changes are well reproduced by convolution of the theoretical cross-sections with experimental low-loss spectra (compare Fig.~\ref{fig_13}b
and Fig.\ref{fig_13}c).

This strongly supports the necessity of recording low-loss spectra alongside core-loss spectra for accurate quantification. In STEM-EELS, a good practice is to employ Dual-EELS systems that simultaneously acquire low-loss and core-loss regions or hybrid cameras capable of capturing both energy regions in a single acquisition.

\subsection{Approximate Correction in the Absence of Low-Loss Spectra}

However, in many cases low-loss spectra cannot be acquired due to instrumental or time limitations. In that cases, plural scattering correction has typically been omitted, potentially leading to significant quantification errors.

Is it possible to correct for plural scattering \textit{approximately}, based on few information about instrumentation and a sample eventually known \textit{a priori}? Answering this question requires extensive  modeling to determine which features of the low-loss spectrum most significantly influence the shape of the core-loss signal.

The simplest correction involves accounting for the energy resolution of the spectrometer, which is usually known for a given instrument setting. The zero-loss peak can be modeled as a Gaussian with the Full Width Half Maximum (FWHM) representing instrumental resolution.

\begin{figure}[ht!]
	\centering
	\caption{Theoretical Cu $L_{23}$ cross-section convolved with a zero-loss peak of different energy resolutions.  For easier comparison with the other edges, the energy-loss is counted from the Cu $L_{23}$ onset (931eV).  The unconvolved cross-section is  dotted. }
	\includegraphics[width=0.5\textwidth]{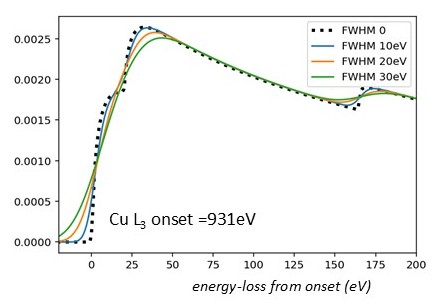}
	\label{fig14}
\end{figure}

Fig.~\ref{fig14} shows the effect of smearing the theoretical cross-section with this Gaussian, demonstrating that the effect is negligible, even for impractically large FWHM values.

As a next step, we simulate plasmon peaks in low-loss spectra using the Drude model that predicts the Lorenzian shape of a plasmon peak (equation (3.43) in \cite{Egerton}):
\begin{equation}
	P = \frac{E W_p E_p^2}{(E^2 -E_p^2)^2 + (E W_p)^2}
	\tag{6}
\end{equation}
where $E$ is energy-loss, $E_p$ is plasmon energy and $W_p$ is a full width at half-maximum of a plasmon peak. Following Egerton (equation (3.94) in \cite{Egerton}), probability of  scattering $P_n$ from multiple independent inelastic events is modeled using Poisson statistics:
\begin{equation}
	P_n = \frac{1}{n!} th^n e^{-th}
	\tag{7}
\end{equation}

\begin{figure}[ht!]
	\caption{(a) A simulated low-loss spectrum for Si for different values of relative thickness $th$ and its influence on the theoretical $K$ edge cross-section. Additionally, the unconvolved cross-section is shown dotted.  (c) shows the experimental spectra and correspondingly, (d) their influence on the cross-section. The energy resolution (FWHM) is 3 eV.  All low-loss spectra are normalized to the height of a zero-loss peak while core-loss spectra are normalized to the total area. The energy-loss in (b,d) is counted from the Si $K$ onset (1839eV).}
	\includegraphics[width=0.99\textwidth]{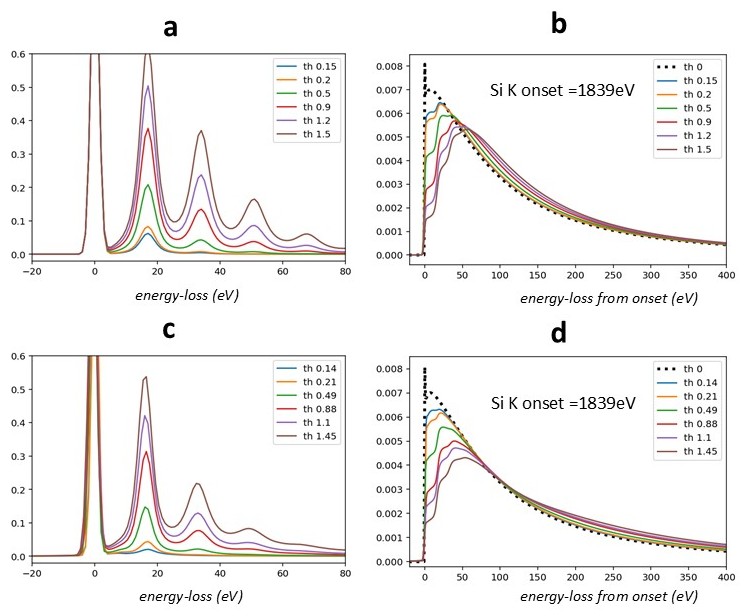}
	\label{fig13}
\end{figure}

Here,  $P_n$ represent the $n$-th order plasmon and $th$ is, as previously, a relative thickness. Apparently, the position of each n-th order plasmon is $E_p$-shifted to the higher energy relative that of the previous order. The probability of higher-order plasmon excitations drops rapidly, thus summing above 7 orders of plasmon excitations does not affect the sum curve.

Fig.~\ref{fig13}a shows a simulation using 7 orders of plasmons in the Drude model for pure Si.
Comparison with experimental low-loss spectra (Fig.~\ref{fig13}c) reveals a reasonable agreement although, in simulations, the inelastic scattering is more concentrated near the lower energies. That is because the  single-electron excitations were not accounted for in the model. Correspondingly, the convolution of the theoretical cross-sections using the simulated low-loss spectra slightly differ from that using the experimental ones (compare Fig.~\ref{fig13}b and d). Namely, it fits well at energies less than 50-100 eV while  deviates at higher energies.

The reason for this discrepancy is a fundamental inability to explain the shape of  low-loss EELS spectra only with plasmon peaks. Indeed, a number of single-particle valence band and intra-band transitions are expected in the low-loss region regardless of whether they are seen explicitly or masked by the plasmon peaks.

In attempt to further improve the accuracy of treatment, we can empirically account for low-energy single particle transitions (Appendix B). However, in most practical cases, a simple Drude model already provides a sufficient approximate correction.

Except for the relative thickness \textit{th}, our empirical correction involves two tuning parameters – a mean plasmon energy  $E_p$ and a plasmon peak width $W_p$. Appendix C demonstrates that the final results are not highly sensitive to the variation of these parameters in the realistic range. Thus, we can fix them at some average  for common materials value: $E_p$ = 20eV, $W_p$ =15eV. The only critically important free parameter the user must estimate for a given TEM sample is its relative thickness $th$.

If there are no hints about the thickness of a sample, the assumption of a \textit{typical} thickness  in TEM (0.4-0.8) would nevertheless give better results than ignoring plural scattering, i.e. assuming de-facto zero sample thickness.

\section{Elements Quantification}
\subsection{Relative Quantification}

The fitting procedure assigns a cross-sectional contribution $p_i$ to each element $i$ included in the quantification. These $p_i$ values are proportional to the number of atoms of a given type contributing to the scattering.

Assuming that all elements present in the sample are included in the quantification, their relative atomic fractions can be obtained as
\begin{equation} \label{eq6}
	a_i = \frac{p_i}{\sum_{j} p_j}
	\tag{8}
\end{equation}

The mass fractions are then calculated as
\begin{equation} \label{eq7}
	m_i = \frac{a_i M_i}{\sum_{j} a_j M_j}
	\tag{9}
\end{equation}
where $M_i$ is the atomic mass of element $i$.

Relative quantification provides a transparent result and does not require knowledge of experimental parameters such as probe size, sample thickness, or beam current.

In some situations, however, this approach may yield inappropriate results.
Spectrum-imaging over regions with strongly varying sample thickness can lead to large variation in the precision of the deduced relative composition. Near holes in samples, the relative fractions may even diverge.

A spectrum averaged over all pixels of a spectrum-image can help mitigate such divergence. If the total spectral intensity in certain pixels falls below a predefined fraction of the average spectrum intensity, the composition in these pixels can be set to zero. This procedure is useful in identification of regions of vacuum or near-vacuum in the sample.

\subsection{Absolute Quantification}

When a spectrum does not contain the edges of all elements present in the sample, relative quantification becomes less effective. In such cases, only ratios between the detected elements can be determined. Furthermore, relative quantification is of no use when the distribution of a single element is required.

In these situations, the absolute elemental content expressed in meaningful physical units is preferred.

Suppose that a sample with area $A$ is irradiated by an electron flux $F$ (in $\frac{\mathrm{electrons}}{\mathrm{seconds}}$) for a time $t$, producing an EELS spectrum. The total signal  $I_i$ from inelastic scattering on atoms of type $i$  is expressed as

\begin{equation} \label{eq8}
	I_i  =  Ft  \frac{N_i}{A} \int \sigma_i(E)=  \frac{1}{\epsilon} ( p_i \int \sigma_i(E)\, dE )
	\tag{10}
\end{equation}

where  $N_i$ is a number of $i$-th atoms in a sample of area $A$ and $\sigma_i(E)$ is the differential cross-section, i.e. effective scattering area for one atom of $i$ kind. The right part of (10) points to the fact that the total recorded signal profile is fitted to $\sigma_i(E)$ through the contribution coefficient $p_i$.  There, we convert the signal from counts to the number of electrons via  the conversion efficiency  of the detector $\epsilon$  in units $\frac{\mathrm{counts}}{\mathrm{electron}}$.

From (10), $p_i$ is expressed as a combination of the density of atoms of type $i$ per unit area,  the integrated electron flux and the conversion efficiency:

\begin{equation} \label{eq9}
	p_i = \epsilon \frac{N_i}{A}\, F t
	\tag{11}
\end{equation}

In most cases, the electron flux is not accurately known. However, it can be indirectly estimated from an  EELS spectrum acquired in vacuum under identical experimental conditions and acquisition time:

\begin{equation} \label{eq10}
	F t = \frac{1}{\epsilon} \int Z(E)\, dE ,
	\tag{12}
\end{equation}

where $Z(E)$ is the energy-dependent intensity of zero-loss peak as recorded by the detector in vacuum.

Combining Eqs.~(11) and (12), the areal density of atoms of type $i$ becomes

\begin{equation} \label{eq11}
	\frac{N_i}{A} = \frac{p_i}{\int Z(E)\, dE} .
	\tag{13}
\end{equation}

The dimensionality of an integrated cross-section is $\left[\frac{\mathrm{barn}}{\mathrm{atom}}\right]$, thus $p_i$ in (10) is expressed in $\left[\frac{\mathrm{atoms}\times \mathrm{counts}}{\mathrm{barn}}\right]$. Since one barn equals $10^{-28}\,\mathrm{m}^2$, Eq.~(13) can be conveniently rewriten as

\begin{equation} \label{eq12}
	\frac{N_i}{A}
	= \frac{p_i}{\int Z(E)\, dE}\times 10^{9}
	\;\; \left[\frac{\mathrm{atoms}}{\mathrm{nm}^2}\right].
	\tag{14}
\end{equation}

In practice, the total intensity of zero-loss peak  is evaluated by summation of the detector counts $Z_i$ over all available energy channels $j$, thus (14) becomes

\begin{equation} \label{eq13}
	\frac{N_i}{A}
	= \frac{p_i}{\sum_j Z_j}\times 10^{9}
	\;\; \left[\frac{\mathrm{atoms}}{\mathrm{nm}^2}\right].
	\tag{15}
\end{equation}

These derivations show that the absolute areal density of atoms of a given element can be obtained from the spectral fit without explicit knowledge of either the electron flux, the probe size or conversion efficiency.

\begin{figure}[ht!]
	\caption{(a): Semiconductor sample with drastically changing thickness from left to right. The measured purely Cu areas with the relative thickness $th$ ranging from 0.10 to 1.76 are marked by red squares. (b) and (c): The Cu $L_{2,3}$ edges from the areas of minimal and maximal thickness $th$. The original spectra are shown in blue, fitted profiles in orange and fitted background in green. (d): The fitted fraction of Cu in different areas plotted as a function of their relative thickness $th$. (e) The integrated area under the complete spectrum in different areas.  (f) The number of Cu atoms per $nm^{2}$  calculated with normalization on the zero-loss peak in vacuum (13) and on the total spectrum counts (14). (b), (c) have a common abscissa - energy-loss; (d)-(f) have a common abscissa - relative thickness $th$. }
	\includegraphics[width=\textwidth]{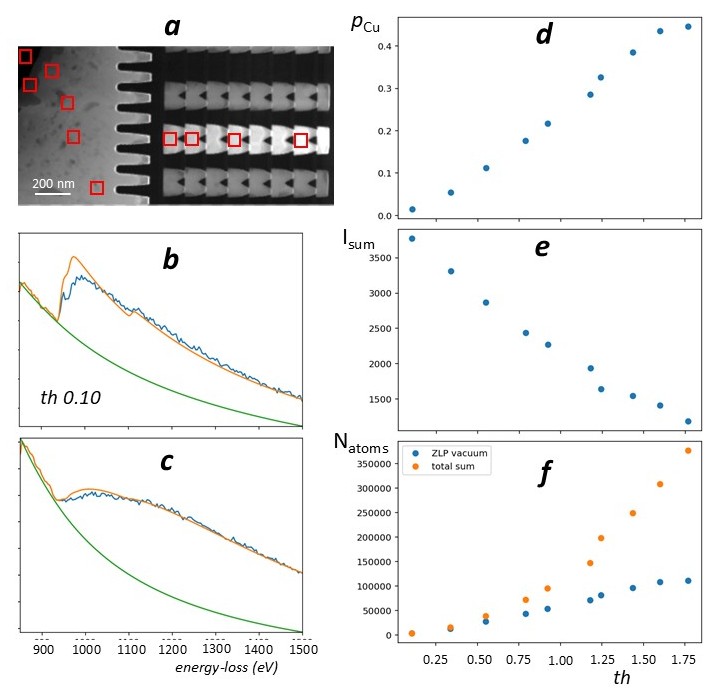}
	\label{fig18}
\end{figure}

A natural question arises: is it preferable to estimate the electron flux in Eq.~(12) using the same spectrum, rather than that acquired in vacuum? A simple approach would be to integrate the entire spectrum over all energy-losses, assuming that all transmitted and scattered electrons are  collected by the detector. Apparently, the experimental spectrum must contain as large energy range as possible \textit{including} a zero-loss peak. Then, formula (13) should be replaced with:

\begin{equation} \label{eq14}
	\frac{N_i}{A}
	= \frac{p_i}{\sum_j S_j}\times 10^{9}
	\;\; \left[\frac{\mathrm{atoms}}{\mathrm{nm}^2}\right].
	\tag{16}
\end{equation}
where $S_j$ the  spectrum recorded over the all available energy channels.

To clarify this question, we performed measurements in the Cu regions of varying thickness, while keeping all experimental conditions and the electron flux strictly identical (Fig.\ref{fig18}a). Figs.\ref{fig18}b and \ref{fig18}c show that, despite substantial changes in plural scattering, the calculations reproduce the experimental signal profiles quite well. As expected, the fitted copper contribution $p_{Cu}$ increases nearly linearly with increasing sample thickness (Fig.\ref{fig18}d).

However, the total spectral intensity, obtained by integrating the zero-loss peak together with all accessible energy-losses, is not constant. Instead, it decreases   with increasing sample thickness, as shown in Fig.\ref{fig18}e. Attempts to extrapolate the spectra to quasi-infinite energy-loss using the power-law tails did not noticeably improve the situation.

This  behavior is most likely the consequence of elastic scattering, which, in contrast to inelastic one, extends to very large scattering angle \cite{Egerton} and therefore falls outside the detector acceptance angle. Consequently, the accurate measurement of the electron flux from the  total counts is questioned.

On the other hand, the part of the Cu $L_{2.3}$ signal should also be elastically scattered outside the detector in the proportion comparable to that for the incident electrons. Thus, these two effects could compensate each other. Such an assumption is however not evident as the core-loss signal is incoherent while the rest spectrum has a mixed, coherent and incoherent nature, therefore their scattering might differ.

Fig.\ref{fig18}f shows a number of Cu atoms per $nm^2$ calculated with both kinds of normalization (on the vacuum zero-loss peak  (15) and on the total spectrum counts  (16)). The results are very similar for thin samples but diverge at the larger sample thickness.

The parameter $th$ calculated from the low-loss spectra  (equation (3.95) in \cite{Egerton}) is believed to accurately track the relative thickness of a sample. Thus, the dependence in Fig.\ref{fig18}f is expected to be linear. However, at large thicknesses, the curve normalized as  (16) deviates upwards, while the curve normalized as  (15) deviates slightly downwards from the expected linearity. This might indicate that the former normalization method overestimates the number of atoms while the later one underestimates it. For the moment, there is no clear experimental evidence in favor of either this or that method of normalization; thus, further studies are desired.


\section{Concluding Remarks}

The various issues in EELS quantification, as well as the possible solutions reviewed above, demonstrate that even after more than 50 years of development, this field is still far from a “nothing-to-improve” state. Progress in both EELS instrumentation and data-analysis algorithms continues at an undiminished pace. Although general trends and some established solutions are well recognized, new approaches still need to be explored and critically evaluated.

\section{Acknowledgment}

The authors acknowledge the consultation, manuscript reviewing and valuable remarks of Prof. Helmut Kohl, University of Münster. Pavel Potapov acknowledges the financial support of CEOS GmbH.

\newpage
\appendix
\section*{Appendix A}

In this appendix, we examine smooth functions \( y(x) \) that are appropriate for modeling background trends on a logarithmic scale. We start with a linear descending relationship between two given points \( x_0 \) and \( x_1 \), and consider a family of functions \( y(x) \) that slightly deviate from this linear trend.

The boundary conditions require that \( y(x_0) \) and its derivative \( y'(x_0) \) match those of the linear function, while \( y(x_1) \) and \( y'(x_1) \) may deviate. To ensure smoothness, we define \( y(x) \) as a cubic polynomial connecting  points $(x_0, y_0)$ and  $(x_1, y_1)$ with the parameters defined by selecting values for \( y(x_1) \) and \( y'(x_1) \).

The objective is to find constraints on the boundary values \( y(x_1) \) and \( y'(x_1) \) such that \( y(x) \) i) is monotonically decreasing and ii) does not exhibit pronounced convex shoulders between \( x_0 \) and \( x_1 \). 

\begin{figure}[ht!] \label{figA1}
	\caption{Schematic plot of linearly descending function and cubic polynomials deviating from that to the positive or negative side. x and y are dimensionless coordinates.}
	\centering
	\includegraphics[width=0.8\textwidth]{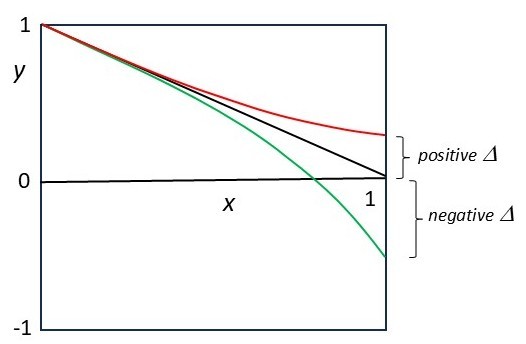}
    \label{fig17}
\end{figure}

The analysis is simplified using dimensionless coordinates: let \( x_0 = 0 \), \( y_0 = 1 \), \( x_1 = 1 \), and \( y_1 = 0 \), as shown in Fig.\ref{fig17}.

For the family of polynomials \( y(x) \), we define:
\[
	y(1) = \Delta, \quad y'(1) = \Theta.
\]
To simplify the derivation, we couple \( \Theta \) and \( \Delta \) through a new non-negative parameter \( k \):
\[
	\Theta = -1 + k\Delta.
\]

The polynomial has the general form:
\begin{equation} \label{eqA1}
	y(x) = ax^3 + bx^2 + cx + d
	\tag{A1}
\end{equation}

with first and second derivatives:
\begin{equation} \label{eqA2}
	y'(x) = 3ax^2 + 2bx + c
	\tag{A2}
\end{equation}

\begin{equation} \label{eqA3}
	y''(x) = 6ax + 2b
	\tag{A3}
\end{equation}

Given \( y(0) = 1 \) and \( y'(0) = -1 \), we find \( d = 1 \) and \( c = -1 \). For \( y(1) = \Delta \) and \( y'(1) = -1 + k\Delta \), we obtain the system:
\begin{align}
	a + b   & = \Delta
	\tag{A4}            \\
	3a + 2b & = k\Delta
	\tag{A5}
\end{align}

Solving, we find:
\[
	a = \Delta(k - 2), \quad b = \Delta(3 - k).
\]

The polynomials and its derivatives then become:
\begin{equation} \label{eqA6}
	y(x) = \Delta(k - 2)x^3 + \Delta(3 - k)x^2 - x + 1
	\tag{A6}
\end{equation}

\begin{equation} \label{eqA7}
	y'(x) = 3\Delta(k - 2)x^2 + 2\Delta(3 - k)x - 1
	\tag{A7}
\end{equation}

\begin{equation} \label{eqA8}
	y''(x) = 6\Delta(k - 2)x + 2\Delta(3 - k)
	\tag{A8}
\end{equation}

\begin{figure}[ht!] \label{figA2}
	\caption{Point $x^*$ where the second derivative is zero  as a function of k. The regions when intersection is between 0 and 1 are plotted in red}
	\centering
	\includegraphics[width=0.7\textwidth]{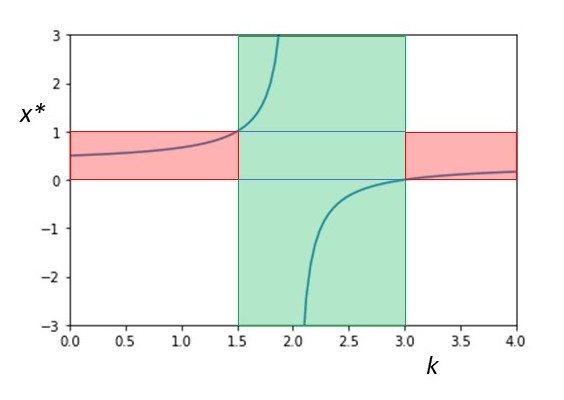}
    \label{fig17b}
\end{figure}

\subsection*{Monotonicity}

For monotonic decrease, \( y'(x) < 0 \) for all \( x \in [0,1] \). At \( x = 0 \), the derivative is always negative. At \( x = 1 \), we require:
\begin{equation} \label{eqA9}
	k\Delta < 1
	\tag{A9}
\end{equation}

To avoid the case where \( y'(x) \) becomes positive within \( (0,1) \), we require that \( y''(x) \neq 0 \) in \( (0,1) \) \footnote{Here we denote the range \textit{inclusive} the endpoints as [] and one \textit{exclusive} the endpoints as ().}. Setting \( y''(x^*) = 0 \) gives:
\begin{equation}
	6\Delta(k - 2)x^* + 2\Delta(3 - k) = 0 \quad \Rightarrow \quad x^* = \frac{1}{3} \cdot \frac{k - 3}{k - 2}
	\tag{A10}
\end{equation}

From Fig.\ref{fig17b}, \( x^* \in (0,1) \) when \( k < \frac{3}{2} \) or \( k > 3 \). Therefore, a sufficient condition to maintain monotonicity throughout \( (0,1) \) is:
\begin{equation}
	\frac{3}{2} < k < 3
	\tag{A11}
\end{equation}

Combining with Eq.~\eqref{eqA9}, we obtain the constraint for $\Delta$:
\begin{equation}
	\Delta < \frac{1}{3}
	\tag{A12}
\end{equation}

\subsection*{Convexity}

\paragraph{Case \( \Delta \geq 0 \):}

Convexity is conserved if the second derivative \( y''(x) \) is positive for all \( x \in [0,1] \). At \( x = 0 \):
\begin{equation}
	2\Delta(3 - k) > 0 \quad \Rightarrow \quad k < 3
	\tag{A13}
\end{equation}

At \( x = 1 \):
\begin{equation}
	6\Delta(k - 2) + 2\Delta(3 - k) > 0 \quad \Rightarrow \quad k > \frac{3}{2}
	\tag{A14}
\end{equation}

Furthermore, it is easy to see that the second derivative \( y''(x) \) has no extrema within \( (0,1) \).

\begin{figure}[ht!] \label{figA3}
	\caption{Second derivative $y''$ in case of \( \Delta < 0 \) for different k. The region of discontinuity is marked in red.}
	\centering
	\includegraphics[width=0.6\textwidth]{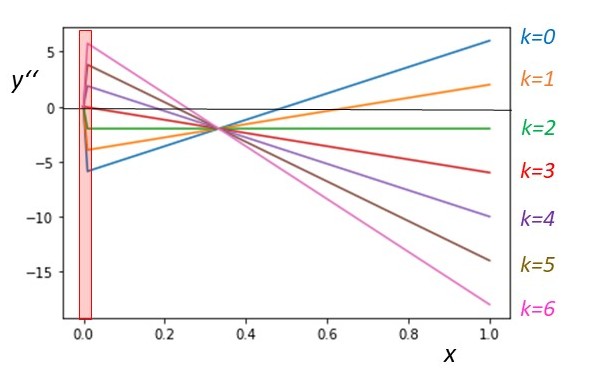}
    \label{fig19_}
\end{figure}

\paragraph{Case \( \Delta < 0 \):}

As seen from Fig.\ref{fig17}, the spline is not convex in this case; there is always a point above the line connecting \( y(0) \) and \( y(1) \). To avoid pronounced nonconvex shoulders, we aim to minimize the deviation of the second derivative towards the negative side and avoid its sharp variation.

\begin{figure}[ht!]
	\caption{(a) Simulation of Cu low-loss spectra with the Drude plasmon model and (b) corresponding convolution of the $L_{2,3}$ theoretical cross-sections. The plasmon energy was assumed to be 20 eV with the peak width 25 eV.  (c) and (d)  show the corresponding results with the augmented Power Law tails of $r=1.5$. The experimental low-loss and core-loss spectra are shown in (e, f).   }
	\includegraphics[width=0.92\textwidth]{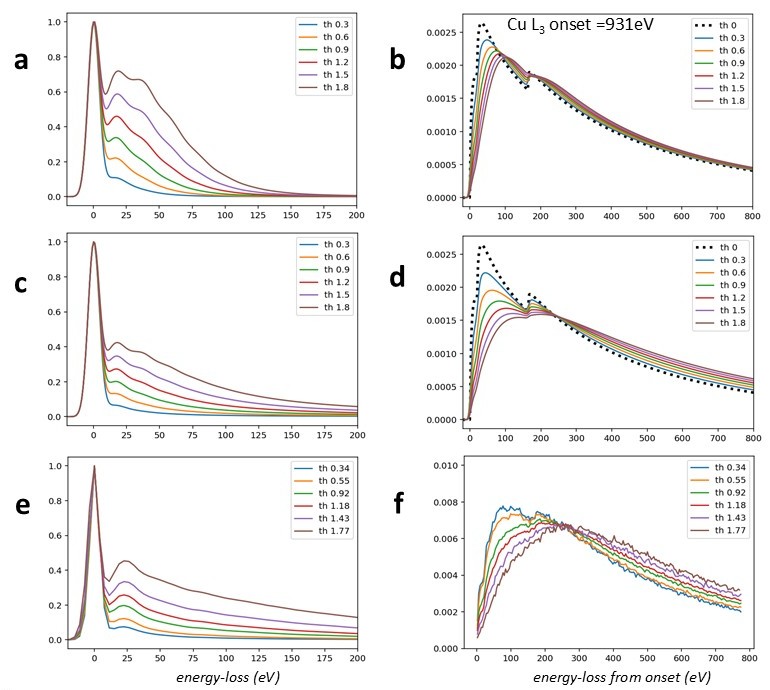}
	\label{fig_20}
\end{figure}

Fig.\ref{fig19_} plotting \( y''(x) \) for various \( k \) also highlights another issue. Because the curve must fit the linear function at the left end, the second derivative has a discontinuity at \( x = 0 \), except when \( k = 3 \). 

For \( k = 2 \), the second derivative is constant throughout the considered range. Hence, choosing \( k \approx 2 \) uniformly distributes the curvature. Meanwhile, \( k \approx 3 \) avoids peaked behavior, but for \( k > 3 \), the second derivative falls sharply into the negative range at the right side.

Taking into account these trade-offs, a reasonable compromise is to choose
\[
	\frac{3}{2} < k < 3,
\]
which coincides with the condition for convexity at $\Delta >0$ and the condition for monotonity.

\newpage
\appendix
\section*{Appendix B}

Simulation of low-loss spectra with the Drude plasmon model might be insufficient in cases where material shows a number of single particle transitions at low energies. We investigate this case on example of pure Cu that is known to exhibit a number of valence  and intra-band transitions as well as the $M$ core excitations in the low-loss region.

Fig.\ref{fig_20}a shows the simulation of low-loss spectra with the simple Drude plasmon model that produces the spectra largely concentrated near the zero-loss peak. This does not reproduce well the experimental extended tails towards higher energies (Fig.\ref{fig_20}e). The experimental data are copied from Fig.\ref{fig_13}a,b and displayed here for the easy comparison.  Consequently, the convolved theoretical cross-sections Fig.\ref{fig_20}b differ significantly from the experimental ones in Fig.\ref{fig_20}f.

\begin{figure}[ht!]
	\caption{The plasmon peak (blue) calculated in the Drude model with a plasmon energy of 20 eV and width of 25eV. The tail of the plasmon can be extended by augmentation with the empirical Power Law with different exponents (shown in red, green, orange). }
	\includegraphics[width=0.5\textwidth]{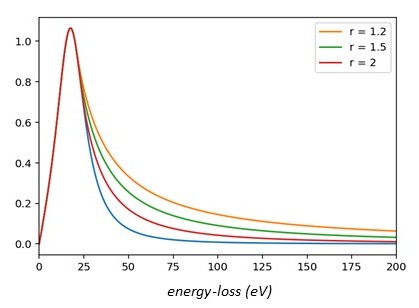}
	\label{fig16}
\end{figure}

Neither the Lorentz-oscillator nor Mermin extensions of the Drude model  improve the match noticeably. To address the difference, we augment the Drude plasmon peak with a slowly decaying empirical power-law tail as illustrated  in Fig.\ref{fig16}. This aims to mimic a tail from single-particle excitations located \textit{approximately} at the same energy and masked by the intensive plasmon peak. The power-law exponent $r$, may range from 1 to 3 in order to match the experimental observations. The augmentation point is defined where the derivatives of the Drude and Power Law functions coincide. We emphasize that this empirical approach does not aim  to reproduce real low-loss spectra (they still differ noticeably), but rather to correct roughly for  the effect of unknown excitations near the plasmon energy and therefore make the results of convolution more realistic.

Fig.\ref{fig_20}c shows the calculations with such an augmented Drude model where plural scattering was again generated by formula (4). The resulting low-loss tails are more prolonged and qualitatively consistent with experimental ones. Accordingly, the convolved theoretical core-loss cross-sections (Fig.~\ref{fig_20}d) align more closely with experimental ones (Fig.~\ref{fig_20}f).

\newpage
\appendix
\section*{Appendix C}

\begin{figure}[ht!]
	\caption{Theoretical O $K$ cross-sections convolved with the simulated low-loss for varying sample thickness  up to $th$=1.8. The tableau shows the results for different pseudo-plasmon energies $E_p$ (different columns) and plasmon width $W_p$ (different rows). For easier comparison with the other edges, the energy-loss is counted from the O $K$ onset (532 eV).}
	\centering
	\includegraphics[width=\textwidth]{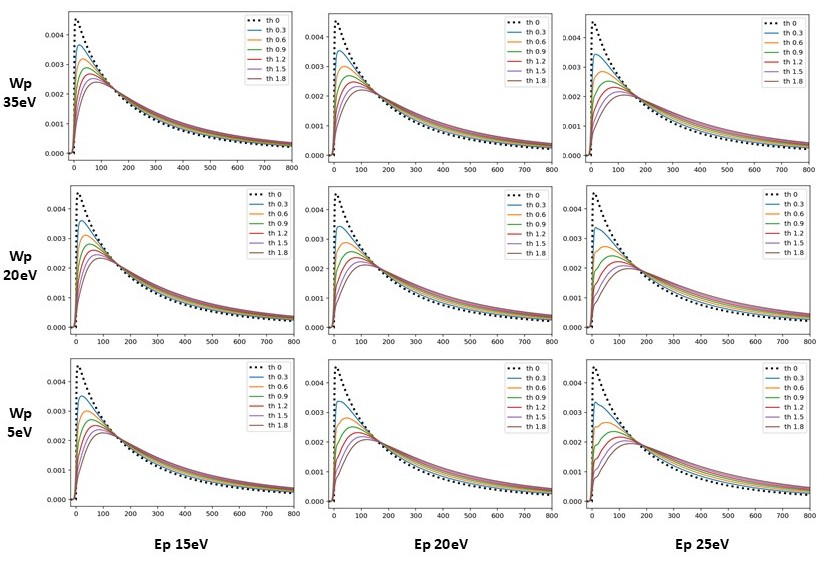}
	\label{figC1}
\end{figure}
\begin{figure}[ht!]
	\caption{Theoretical Si $K$ cross-sections convolved with the simulated low-loss for varying sample thickness up to $th$=1.8. The tableau shows the results for different pseudo-plasmon energies $E_p$ (different columns) and plasmon width $W_p$ (different rows). For easier comparison with the other edges, the energy-loss is counted from the Si $K$ onset (1839 eV)}
	\centering
	\includegraphics[width=\textwidth]{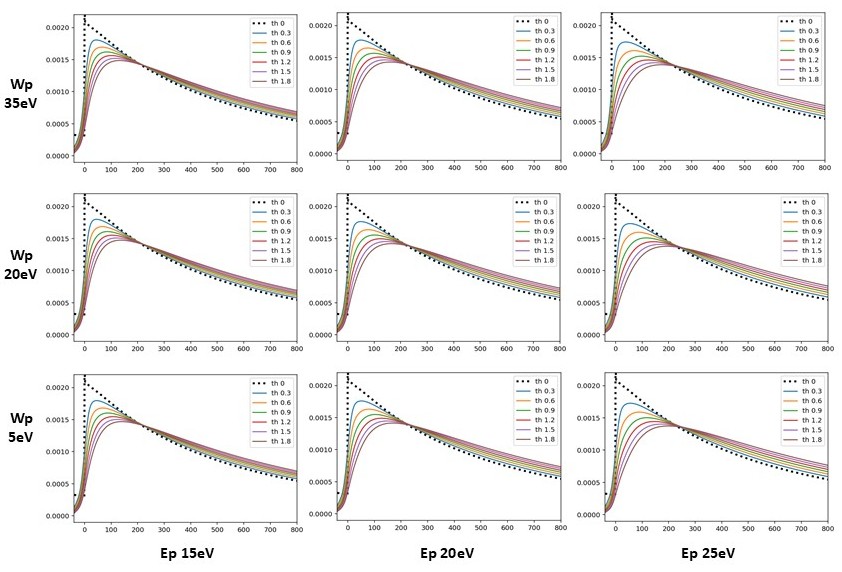}
	\label{figC2}
\end{figure}

The sensitivity of approximate convolution of the theoretical cross-sections to the modeling parameters was examined on two typical EELS edges - O $K$ and Si $K$. A plasmon in the Drude model is characterized by the plasmon energy $E_p$ and the plasmon peak width $W_p$. Fig.\ref{figC1} and \ref{figC2} show the empirical convolution of O $K$ and Si $K$ cross-sections for different $E_p$ and $W_p$ values. The results show minimal sensitivity to $W_p$, and only a slight dependence on $E_p$.

\newpage

\bibliographystyle{unsrt}
\bibliography{references}

\end{document}